**Thermally driven escape from Pluto's atmosphere: A combined fluid/kinetic model**

O. J. Tucker[*], J. T. Erwin, J. I. Deighan, A. N. Volkov, R. E. Johnson

Engineering Physics, 395 McCormick Road, University of Virginia, Charlottesville, Va, 22904 USA

Email Addresses: ojt9j@virginia.edu [*], jte2c@virginia.edu, jid7v@virginia.edu, av4h@virginia.edu, rej@virginia.edu

Phone Number: 1-434-924-3244

**Abstract:** A combined fluid/kinetic model is developed to calculate thermally driven escape of $N_2$ from Pluto's atmosphere for two solar heating conditions: no heating above 1450 km and solar minimum heating conditions. In the combined model, one-dimensional fluid equations are applied for the dense part of the atmosphere, while the exobase region is described by a kinetic model and calculated by the direct simulation Monte Carlo method. Fluid and kinetic parts of the model are iteratively solved in order to maintain constant total mass and energy fluxes through the simulation region. Although the atmosphere was found to be highly extended, with an exobase altitude at ~6000 km at solar minimum, the outflow remained subsonic and the escape rate was within a factor of two of the Jeans rate for the exobase temperatures determined. This picture is drastically different from recent predictions obtained solely using a fluid model which, in itself, requires assumptions about atmospheric density, flow velocity and energy flux carried away by escaping molecules at infinity. Gas temperature, density, velocity and heat flux versus radial distance are consistent between the hydrodynamic and kinetic model up to the exobase, only when the energy flux across the lower boundary and escape rate used to solve the hydrodynamic equations is obtained from the kinetic model. This limits the applicability of fluid models to atmospheric escape problems. Finally, the recent discovery of CO at high altitudes, the effect of Charon and the conditions at the New Horizon encounter are briefly considered.

# 1 INTRODUCTION

Our understanding of Pluto's $N_2$ dominated atmosphere is largely based on rare occultation observations in 1988 (pre-perihelion), 2002 and 2006 (post-perihelion) (Elliot et al., 2007; Young et al., 2008). Each observation determined the surface pressure to be between 6.5-24 µbars with a peak atmospheric temperature of ~100 K at a radial distance ~1250 km. At this temperature and due to Pluto's low gravitational energy, e.g. 0.007 eV/amu at 1250 km, escape can result in significant atmospheric loss. Therefore, many studies of Pluto's evolution have been aimed at understanding the loss rate over time (Hubbard et al., 1990; Hunten and Watson, 1982; Strobel, 2008a; Tian and Toon, 2005; Trafton, 1980).

Using results from the occultation observations and calculations of solar heating rates in continuum models of the upper atmosphere, a series of authors calculated thermal escape rates from Pluto's atmosphere by considering a process referred to as slow hydrodynamic escape (SHE, e.g., McNutt, 1989; Krasnopolsky, 1999; Strobel, 2008a). The SHE model of the atmosphere is based on the premise that thermal energy of molecules is efficiently converted into bulk flow energy. This assumption can lead to an over-estimate of the escape rate (Johnson, 2010; Volkov et al., 2011a, b) when applied to the rarefied region of the atmosphere where the collisions are too infrequent to maintain local thermal equilibrium. Using a combined fluid/kinetic model to directly account for the non-equilibrium nature of the gas flow in the upper atmosphere, thermally driven escape from Pluto's atmosphere is found to occur on a molecule by molecule basis resulting in an enhanced Jeans escape rate (Tucker et al., 2011).

Globally averaged escape rates typically have been estimated using fluid (Strobel, 2008a, b) or kinetic models (Tucker and Johnson, 2009; Volkov et al., 2011a, b). Kinetic models can in principle be applied to the entire atmosphere, but they are computationally expensive when applied to a dense region of the atmosphere. In the dense region of the atmosphere where collisions are frequent the flow can be treated as a continuum and the use of a fluid model is most efficient. In the fluid models the hydrodynamic equations are solved to obtain the mass flow rate through the atmosphere but are unable to account for relatively infrequent collisions of upwardly moving molecules or returning molecules that regulate escape in the upper atmosphere. Therefore, they cannot correctly calculate the amount of heat transported through the atmosphere. A combined fluid/kinetic model applied to dense and rarefied parts of the atmosphere respectively avoids these difficulties and provides a computationally-efficient tool for simulation of the atmosphere (e.g. Marconi et al., 1996).

The aim of the present paper is to obtain the globally averaged escape rate, gas density, temperature, velocity, and the heat flux in Pluto's atmosphere in the region between 1450 km – 30000 km using a combined fluid/kinetic model. We numerically solve the one-dimensional (1D) hydrodynamic equations coupled to a molecular kinetic model for the rarefied region of the atmosphere similar to the approach used in Marconi et al. (1996). Preliminary results were given in Tucker et al. (2011). Here simulations are performed for no heating and solar minimum heating conditions. The interaction of Pluto's extended atmosphere with Charon and the implications of the recently discovered CO detection in Pluto's extended atmosphere (Greaves et al., 2011) are briefly considered. The results presented here also suggest that the application of the hydrodynamic models to escape from other planetary atmospheres (e.g., Murray-Clay et al., 2009; Tian, 2009; Strobel, 2008a, b) can give incorrect estimates of the macroscopic properties and the escape flux.

**2     Jeans, Hydrodynamic, and Slow Hydrodynamic Escape**

Although escape driven by solar heating is by its nature a three-dimensional (3D) process, for comparison with previous models, the thermal escape problem is formulated here by considering a 1D, globally averaged, steady-state model of the atmosphere, where the gas properties are functions of the radial distance $r$, from the planet center. In this section, the applicability of the fluid model to thermal escape is briefly analyzed. Based on this analysis, a combined fluid/kinetic model is introduced in the next section.

The Jeans parameter, $\lambda(r)$, the ratio of the gravitational energy of a molecule $\Phi_g = GM_p m/r$ to its thermal energy $kT$, is often used to characterize the atmospheric escape rate: i.e., $\lambda(r) = \Phi_g / kT$, with $G$ the gravitational constant, $M_p$ the planet mass, $m$ the molecular mass, $k$ the Boltzmann constant and $T$ the temperature at $r$. In order to escape a planet's gravity a molecule must be directed outward from the planet, have a velocity larger than $v_{esc} = \sqrt{2GM_p/r}$ and have a low probability of colliding with other molecules along its trajectory. The density of planetary atmospheres decreases exponentially with altitude and the most rarefied region is referred to as the exosphere. In this region intermolecular collisions are rare, therefore the Jeans parameter is typically evaluated at the lower boundary of the exosphere which is referred to as the exobase $r_x$. Throughout the paper the subscript "x" will be used to denote the values of all parameters evaluated at $r = r_x$.

The degree of rarefaction of a gas is determined by the local Knudsen number, $Kn = l_c / l_a$, the ratio of the mean free path of the molecules, $l_c \sim c/(n\sigma)$, to an appropriate length scale for the gas density $l_a$. Here $\sigma$ is the molecular cross section, $n$ is the local number density, and the numerical coefficient $c$ depends on the model of intermolecular collisions, e. g., for hard sphere molecules $c = 1/\sqrt{2}$ (Bird, 1994; Chapman and Cowling, 1970). The appropriate length $l_a$ for planetary atmospheres is usually defined as the distance over which the density decreases by a factor of *1/e* and is called the atmospheric scale height $H = r / \lambda(r)$, so that $Kn = l_c / H$. When $Kn << 0.1$ the atmosphere is relatively dense so that molecules collide frequently. With increasing altitude collisions become less frequent and the exobase altitude is defined to occur where $Kn_x \sim 1$ or $n_x H_x \sigma \sim 1$.

Three regimes of escape are typically characterized using the Jeans parameter. If a planetary atmosphere has a relatively large Jeans parameter at the exobase for the dominant atmospheric species, thermal escape occurs on a molecule by molecule basis similar to evaporation, a process referred to as Jeans escape (Jeans, 1916). In this approximation it is assumed the speed distribution at the exobase is Maxwellian, so that the molecular loss rate is $\varphi_J = \pi r_x^2 n_x \langle v_{th,x} \rangle (1+\lambda_x) \exp(-\lambda_x)$, where $<v_{th,x}> = (8kT_x/\pi m)^{1/2}$ is the mean thermal speed of molecules at the exobase. The concomitant cooling rate, total energy of molecules escaping the
atmosphere per unit time, is $\langle E\varphi \rangle_J = (kT_x)(2 + 1/(1+\lambda_x))\varphi_J$.

Modified Jeans escape rates, accounting for the non-zero gas velocity at the exobase, have also been proposed (e. g. Chamberlain, 1961; Yelle, 2004; Volkov et al., 2011a, b).

At small Jeans parameters the thermal energy of molecules is comparable to or larger than the gravitational binding energy of the planet at the exobase, so the bulk atmosphere can escape as a hydrodynamic outflow (e.g., Öpik, 1963; Hunten, 1973; Volkov et al., 2011a; Gruzinov, 2011). This is often referred to as blow off resulting in escape rates much larger than the Jeans rate. Blow off has been suggested to occur when $\lambda_x <$ ~2 at the exobase altitude or below (Hunten, 1973; Watson et al., 1981).

The slow hydrodynamic escape (SHE) model, considered intermediate to Jeans and hydrodynamic escape regimes, has been suggested to be applicable to a dense tightly bound atmosphere for which the Jeans parameter, $\lambda(r_0)$ is larger than 10, estimated at a radial distance, $r_0$, considered to be in approximate thermal and radiative equilibrium (e.g., Parker, 1964b; Watson et al., 1981). The flow is referred to as slow because near $r_0$ the gravitational energy $\Phi_g$ dominates the thermal energy ($C_p T$), which also dominates the flow energy ($\frac{1}{2}mu^2$) where $C_p$ is the heat capacity per molecule and $u(r)$ is the flow speed. However the flow

177 eventually reaches supersonic speeds above the exobase resulting

178 in escape rates much larger than the Jeans rates. Below we provide

179 a more detailed discussion on the application of hydrodynamic

180 models to the slow hydrodynamic escape regime.

181     In the 1D steady-state hydrodynamic model the continuity

182 equation leads to a constant molecular flow, given here as a flow

183 rate, $\varphi$, vs. radial distance:

184 $$\varphi = 4\pi r^2 n(r) u(r) = 4\pi r_0^2 n(r_0) u(r_0). \qquad (1a)$$

185 The radial momentum equation, in which the viscous term is

186 dropped, is:

187 $$dp/dr = n(d\Phi_g/dr - d(\tfrac{1}{2}mu^2)/dr) \qquad (1b)$$

188 with the gas pressure $p = nkT$. Finally, the corresponding energy

189 equation is:

190 $$\frac{d}{dr}\left\{\varphi(\tfrac{1}{2}mu^2 + C_p T - \Phi_g) - 4\pi r^2 \kappa(T)\frac{dT}{dr}\right\} = 4\pi r^2 Q(r) \qquad (1c)$$

191 where $\kappa(T)$, is the thermal conductivity, $C_p$ is the heat capacity per

192 molecule and $Q(r)$ accounts for the solar heating and IR cooling

193 rates. Knowing the number density, $n_0$, and temperature, $T_0$, at the

194 lower boundary, i.e. $n(r_0) = n_0$, $T(r_0)=T_0$, Eq. 1b and 1c are solved.

195 A unique solution requires two additional parameters at the lower

196 boundary, $u_0$ (or $\varphi$) and $(dT/dr)_0$. Unfortunately, in order to find $u_0$

197 and $(dT/dr)_0$, one needs to impose assumptions about the solution

198 behavior at $r \to \infty$.

Parker (1958) used the hydrodynamic equations to model the thermal expansion of the solar wind in the vicinity of $\lambda(r_0) \sim 2$. He subsequently extended that model to describe the expansion of a stellar wind from a star with a tightly bound corona with $\lambda(r_0) > \sim 10$ for which no or very little heat is deposited above $r_0$ (Parker 1964a, b). In this formulation, escape is powered by the heat flow from below $r_0$ and the conditions imposed are $T, n \rightarrow 0$ as $r \rightarrow \infty$. It was then shown that the dense atmosphere must expand according to a critical solution, where the flow velocity, $u$, gradually increases above the isothermal speed of sound. Purely subsonic solutions were not permitted because they resulted in a finite pressure at infinity.

Chamberlain (1961) re-considered the expansion of the solar wind for subsonic velocities with the condition $T \rightarrow 0$ as $r \rightarrow \infty$. He deemed this approach to be a slow hydrodynamic expansion of the solar wind, and showed it is possible to obtain a subsonic solution with the hydrodynamic equations if the energy flux at infinity is 0. In this formulation the number density $n$ approaches a constant as $r \rightarrow \infty$. Later, Parker (1964b) acknowledged this result as a limiting case to supersonic expansion. He determined that this approximation would only occur in the limit that the density at the lower boundary, $n_0$, goes to infinity. He showed for sufficiently large densities at $r_0$, the energy flux carried to infinity is non-zero for the condition $T \rightarrow 0$ and, hence, the expansion can proceed supersonically. This is the typical approach used in applying the SHE model to planetary atmospheres (e.g., Krasnopolsky, 1999; Strobel, 2008a, b; Watson et. al., 1981). Since the flow is slow, the standard procedure is to integrate Eq. 1b and 1c neglecting the $u^2$ terms (Parker, 1964b). Therefore, although $\varphi \neq 0$ in Eq. 1a, $u^2$ is set equal to 0 in Eq. 1b and 1c below an upper boundary where $\frac{1}{2}mu^2 \ll C_pT$, and $T$ and $n$ are only regarded as valid out to an $r$ where the $u$ is a small fraction of the local sound speed (McNutt, 1989; Krasnopolsky, 1999; Strobel, 2008a, b).

The SHE model has been subsequently applied to the thermal expansion of planetary atmospheres in which solar EUV and UV heating powers escape above $r_0$ (Watson et. al., 1981). Particular emphasis has been placed on Pluto's atmosphere which is widely thought to be escaping hydro-dynamically (e.g., McNutt, 1989; Krasnopolsky, 1999; Strobel, 2008a; Tian and Toon, 2005). For example McNutt (1989) solved the hydrodynamic equations neglecting the term $C_pT$ in Eq. 1c to obtain an analytical solution for $n$, $T$, and the escape rate $\varphi$, assuming the solar heating occurred in a narrow region of the atmosphere. Krasnopolsky (1999) retained the $C_pT$ term, using numerical methods to solve the hydrodynamic equations.

More recently Strobel (2008a, b) applied the SHE model to the atmospheres of Titan and Pluto using more realistic lower boundary conditions at a radial distance where the atmosphere is in approximate radiative equilibrium. He iteratively solved Eq. 1b and 1c using assumed values of $\varphi$ and $(dT/dr)_0$, to find a solution with the right asymptotic behavior and zero total energy flux at $r \rightarrow \infty$ for Pluto (Strobel, 2008a) and matched to available density data for Titan (Strobel, 2008b). Calculated $N_2$ escape rates from Pluto, $\varphi \sim 9.4 \times 10^{26}$ s$^{-1}$, and Titan, $\varphi \sim 1.5 \times 10^{27}$ s$^{-1}$, were several orders of magnitude larger than the Jeans escape rates calculated using the corresponding SHE model exobase densities and temperatures. For example, the most recent SHE model estimate for escape from Pluto's atmosphere is $\sim 10^3$ times the Jeans rate for the suggested atmospheric structure at solar minimum conditions (Strobel, 2008a).

Tucker and Johnson (2009) tested the results for Titan using a kinetic approach and did not obtain large escape rates. In fact, when the temperature in Titan's upper atmosphere was artificially increased so that $\lambda_x \sim 11$, similar to that at Pluto, the escape rate obtained was enhanced over the Jeans rate but only by a factor of ~1.5. Using a kinetic model, Volkov et al. (2011a, b) showed that thermal escape rate from both monatomic and diatomic atmospheres, for which most of the heating occurs below $r_0$, differs from the Jeans rate by less than a factor of 2 if $\lambda(r_0) >\sim 6$. Such a drastic difference between fluid and kinetic simulations in the domain of the slow hydrodynamic escape, $\lambda(r_0) >\sim 10$, is due to the incorrect treatment of the rarefied region of the atmosphere in the hydrodynamic approximation.

The discrepancy between results obtained from solving the hydrodynamic equations and kinetic simulations can be resolved by using a combined fluid and kinetic approach (e.g., Marconi et al., 1996). A stand-alone kinetic simulation is computationally infeasible at $Kn(r_0) \sim 10^{-6}$ characteristic of the density at the lower boundary typically used in modeling escape from Pluto's atmosphere. Therefore, a computationally efficient model can be constructed by coupling the hydrodynamic equations for the dense atmosphere with kinetic simulations for the exosphere region.

### 3     Combined Fluid/Kinetic Model of Thermal Escape

A fluid/kinetic model is applied from a lower boundary $r_0$ in the atmosphere, considered to be in approximate local thermodynamic equilibrium, to a top boundary $r_1$ where the atmospheric flow is essentially free of collisions. We divide the atmosphere into two regions, a fluid region where the hydrodynamic equations are applicable from $r_0$ where $Kn \ll 1$ to an intermediate boundary $r_{od}$ chosen to correspond to $Kn \sim 0.1$, and a kinetic region where kinetic simulations are performed by means

291  of the direct simulation Monte Carlo (DSMC) method (Bird,

292  1994), from $r_{od}$ to $r_1$ where $Kn \gg 1$.

293  When solving the fluid equations we make no assumptions

294  about the $n$ and $T$ at infinity. Consistent with the SHE model we

295  drop the $u(r)^2$ terms. It is possible to include such terms in both the

296  SHE and fluid/kinetic approaches. However, for a dense

297  gravitationally bound atmosphere $u(r)^2$ can be safely neglected

298  below the exobase (Parker, 1964b). The lower boundary conditions

299  in the fluid region are $n_0$ and $T_0$, while the parameters $\varphi$ and $\langle E\varphi \rangle_{r_0}$,

300  the particle and energy flow across $r_0$, in Eq. 2b are determined by

301  the DSMC part of the model. The pressure and the heat flow, given

302  by Eq. (2a, b), are determined from the integration of Eq. 1b and

303  1c using the total heating rate $\beta(r) = r_0^{-2} [\int_{r_0} r^2 Q(r) dr]$ with $\beta \rightarrow \beta_0$

304  as $r \rightarrow \infty$, as defined in Strobel (2008a), and $\langle E\varphi \rangle_{r_0}$ is obtained as

305  the constant of integration:

306  $$p = p_0 \exp\left[-\int \frac{\lambda(r)}{r} dr\right] \quad (2a)$$

307  $$\varphi(C_p T - \Phi_g) - 4\pi r^2 \kappa(T) \frac{dT}{dr} = \langle E\varphi \rangle_{r_0} + 4\pi r_0^2 \beta(r) \quad (2b)$$

308  Since the fluid model requires initial values for $\varphi$ and $\langle E\varphi \rangle_{r_0}$, for

309  the $\beta_0 = 0$ case we began by assuming an isothermal, hydrostatic

310  atmosphere, $d(nkT_0)/dr = n [d\Phi_g/dr]$, and used a DSMC simulation

311  for such an atmosphere starting at $Kn(r_{od}) \sim 0.1$ to obtain the

312    initial estimates. Therefore unlike the SHE model we do not

313    assume that the energy flow at infinity is zero. In a steady state

314    atmosphere energy conservation requires that the energy carried

315    off by escaping molecules is replaced by a flow of energy into the

316    lower boundary, $\langle E\varphi \rangle_{r_0}$, with $\langle E\varphi \rangle = \langle E\varphi \rangle_{r_0} + 4\pi r_0^2 \beta_0$. With such

317    starting conditions, the fluid/kinetic model typically obtained a

318    converged solution in 4 iterations. That is, the temperatures agree

319    within < 3% and densities agree within < 2% in the region where

320    $0.1 < Kn < 1$.

321    As schematically presented in Fig. 1 we solve the Eq. (2a,

322    b) for the density and temperature up to the exobase and iteratively

323    obtain $\varphi$ and $\langle E\varphi \rangle_{r_0}$. From that solution the resulting $n_{od}$ and $T_{od}$ at

324    a radius $r_{od}$ where $Kn \sim 0.1$, which is about two scale heights

325    below the nominal exobase, are used in the DSMC simulation up

326    to an altitude many scale heights above the exobase, $Kn \gg 1$. The

327    DSMC method tracks a representative sample of atmospheric

328    molecules which are under the gravitational influence of Pluto and

329    subjected to mutual collisions. At the upper boundary of the kinetic

330    domain we obtain the particle escape rate, $\varphi$, and the energy flow

331    through the 1D system, $\langle E\varphi \rangle$. Values of $\langle E\varphi \rangle$ and $\varphi$ from the

332    upper boundary are used to update corresponding values in the

333    fluid part of the model and then used to solve Eq. (2a, b) up to

334    $Kn=1$ during the next iteration. Likewise the results from the new

335  simulation of the hydrodynamic equations provide updated $n_{od}$ and
336  $T_{od}$ at $Kn(r_{od}) \sim 0.1$ for the kinetic model, which are then used to
337  obtain new values of $\varphi$ and $\langle E\varphi \rangle$. This procedure is repeated until
338  we reach consistent densities, temperatures and flow velocities in
339  the region where the fluid and kinetic model overlap. That is, for
340  $Kn \leq 1$ we numerically solve Eq. 2a and 2b to obtain $n(r)$ and $T(r)$.
341  The flow velocity, $u(r)$, is given by Eq. 1a for the set of escape
342  parameters $\varphi$ and $\langle E\varphi \rangle$ obtained from the DSMC simulations. We
343  consider a converged solution acceptable when the temperatures
344  and densities agree within $\sim < 3\%$ in the overlap region between
345  our fluid and kinetic models where $0.1 < Kn < 1$.
346

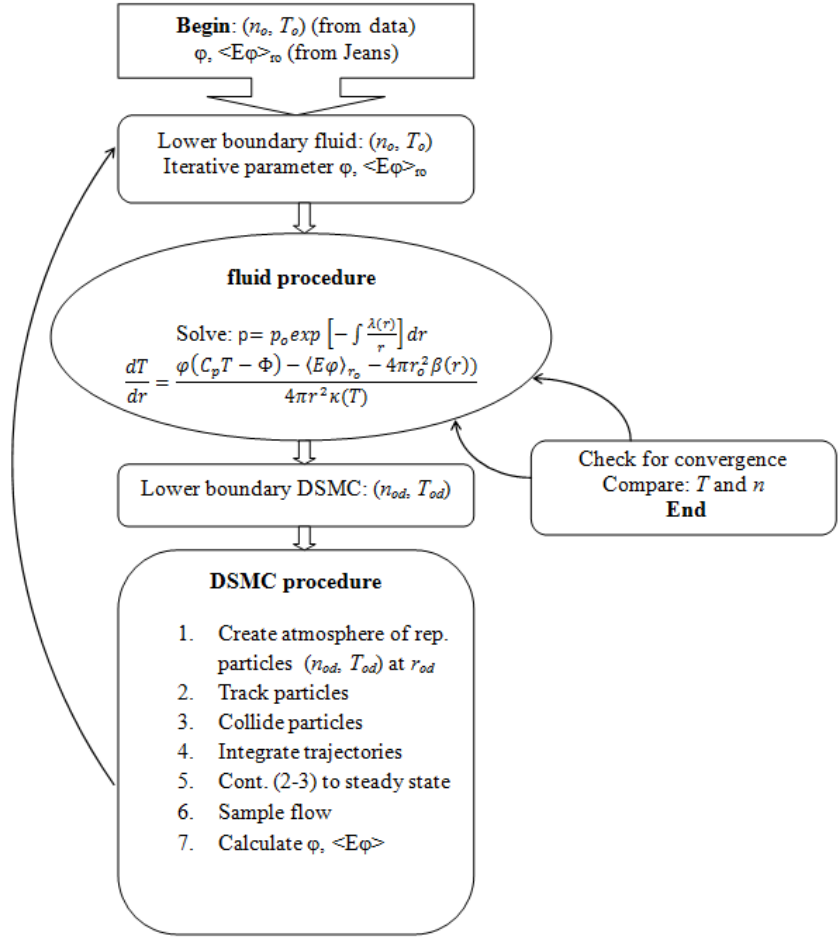

**Figure 1:** Schematic of the numerical implementation of the fluid/kinetic model. To obtain solutions of the hydrodynamic equations at $r_0$ where $Kn < 0.1$, $n_0$ and $T_0$ are given, and $\varphi$ and $\langle E\varphi \rangle_{r_0}$ are obtained iteratively using the DSMC method. An initial guess of the energy flow into the lower boundary $\langle E\varphi \rangle_{r_0}$ is used to solve Eq. 2a and 2b up to $Kn\sim 1$. From the fluid solution $n_{od}$ and $T_{od}$ calculated at $Kn_{od} \sim 0.1$ are used in the DSMC simulation up to $Kn \gg 0.1$ to obtain updated $\varphi$ and $\langle E\varphi \rangle_{r_0}$. The iterations are continued until temperature, density and heat flux are consistent

between the fluid and DSMC solutions in the overlap region of
0.1< $Kn$ <1.

In the low Knudsen number regime Eq. 2a and 2b were solved simultaneously using a $4^{th}$ order Runge-Kutta method with the adaptive radial step-size Runge-Kutta-Fehlberg method (Burden and Faires, 2005), to ensure a relative tolerance of $10^{-8}$ for $n(r)$ and $T(r)$. The integration steps were between 0.1 and 2 km with the finer resolution necessary to resolve the faster change in temperature near the lower boundary and the narrow heating peak. The heating/ cooling models described in Strobel (2008a) are used for the solar heating due to $N_2$ and $CH_4$ absorption bands (including UV, EUV, and near-IR) and CO radiative cooling. Since the heating and cooling rates depend on the temperature and column of gas above a given radial position in the atmosphere, the spatial distribution in the net heating rate is recalculated using the new density profile. These iterations are performed until energy conservation is achieved between $\langle E\varphi \rangle_{r_0}$, $\beta_0$, and $\langle E\varphi \rangle$.

The DSMC method applied in the kinetic region $Kn > 0.1$ in effect solves the Boltzmann kinetic equation by the directly modeling the stochastic nature of the molecular motion in the gas flow using Monte Carlo techniques (Bird, 1994). The DSMC method uses a set of modeling molecules in order to calculate the gas properties of the atmosphere at a molecular level. Collisions between molecules are calculated in discrete radial cells based on the local values of the relative speed, cross section and density. Therefore, the DSMC method is a direct approach for describing the transition in an atmosphere from collisional to collisionless flow. In such a model the conductive heat transfer is represented microscopically.

At the lower boundary of the DSMC domain, $r_{od}$, the density $n_{od}$ and temperature $T_{od}$ are taken from the solutions of the fluid equations as discussed. Although in the DSMC simulations molecular motion and collisions are tracked in 3D, in this paper, Pluto's atmosphere is assumed to be spherically symmetric and so the resulting properties depend only on $r$, consistent with previous models for Pluto's atmosphere (Krasnopolsky, 1999; McNutt, 1989; Strobel, 2008a). Therefore, in the simulation domain when evaluating the collision probabilities the molecular velocities and positions are rotated to a common radial axis. When molecules traverse the DSMC upper boundary, those with velocities greater than the escape velocity and directed outward are assumed to escape and the others are specularly reflected. The reflected molecules represent molecules with trajectories that would eventually return to the simulation domain.

402  In the DSMC method the time step is chosen to be much
403  smaller than the mean collision time and the cell widths in the flow
404  direction are kept much smaller than $l_c$ and $H$ following the general
405  recommendations in Bird (1994). We use variable cell widths in
406  the radial direction which were approximately 1/3$^{rd}$ of the local
407  mean free path and capped at 10% of the local scale height for
408  mean free paths larger than the local atmospheric e-folding.  A
409  time step of ~1 – 2.5s provided energy conservation and ensured
410  that every molecule would have no more than 1 collision over a
411  time step on average. After ~5 x 10$^6$ s the macroscopic properties
412  of the flow were sampled for an additional 5 x 10$^7$ s.  The number
413  of representative molecules was chosen to ensure a sufficient
414  number of molecules (>200) in the upper most cell, typically we
415  used several 10$^3$ -10$^5$ representative molecules to describe the flow
416  in the kinetic region. The upper boundary location was increased
417  until the escape rate varied by less than 5% with increasing upper
418  boundary. Likewise, when using the converged fluid solution with
419  a DSMC lower boundary deeper in the atmosphere, i.e., choosing a
420  point from the fluid solution between $0.01 < Kn(r_{od}) < 0.1$, did not
421  significantly affect the results.
422  Collisions between atmospheric molecules were computed
423  using both the hard sphere (HS) model and the variable hard
424  sphere (VHS) model (Bird, 1994).  To ensure consistency between the fluid and kinetic models, we also used the Larsen-Borgnakke (LB) approximation for internal energy and the VHS cross section is parameterized to the temperature dependent thermal conductivity $\kappa(T) = \kappa_0 T^\omega$ for the Maxwell gas, $\omega = 1$. The VHS cross section, relevant for low speed molecular collisions, depends on the relative collision speed $v_r$, $\sigma = \sigma_0(<v_{r0}>/<v_r>)$: where $\sigma_0$ is a reference cross section determined from the thermal conductivity and $<v_{r0}>$ is the average relative velocity with both values obtained for $T_0$ assuming the Maxwellian speed distribution. At temperatures characteristic for Pluto's upper atmosphere, the $N_2$ vibrational modes are assumed not to be excited so that the LB model is used only for two rotational degrees of freedom. The initial internal energy for each molecule is set at the lower boundary of the DSMC regime, $r_{od}$, based on a Maxwell-Boltzmann energy distribution and neglecting changes in rotational levels due to IR cooling between collisions.

**4    Results for Pluto's Atmosphere**

**Table 1: Parameters for fluid/kinetic model**

| Parameter | HS model | VHS model |
|---|---|---|
| heat capacity/molecule: $C_p$ | $(3/2)k$ | $(5/2)k$ |
| viscosity exponent: $\omega$ ($\kappa(T) = \kappa_0 T^\omega$) | $1/2$ | *1 |
| collision cross section: $\sigma$ (x$10^{-15}$) cm$^{-2}$ | $\sigma_0 = 9.0$ | +$\sigma = \sigma_0(<v_{r0}>/v_r)$ |

Parameters for $N_2$ used in the fluid/kinetic. The lower boundary radial distance is $r_0 = 1450$ km where $n(r_0) = 4 \times 10^{12}$ cm$^{-3}$ ($Kn_0 \sim 10^{-6}$), $T(r_0) = 88.2$ K. *The viscosity exponent for the VHS model and $\kappa_0 = 9.37$ erg cm$^{-1}$ s$^{-2}$ K$^{-2}$ are taken from Strobel (2008a). $^+$The average relative velocity at $r_0$ is defined by $<v_{r0}> = (16kT_0/\pi m)^{1/2}$.

The 1D radial fluid/kinetic model is applied to a region in Pluto's atmosphere from $r_0 = 1450$ km up to $r_1 = 30000$ km for $n_0 = 4 \times 10^{12}$ cm$^{-3}$ ($Kn_0 \sim 10^{-6}$) and $T_0 = 88.2$ K consistent with Strobel (2008a). Pluto's orbital axis is nearly parallel to its orbital plane which will result in the structure of the atmosphere being non-isotropic over the globe, and the amount of solar heating is also variable dependent upon the relative abundances of CH$_4$ and CO present in the atmosphere. However, for the purpose of this study we assume a globally averaged atmosphere and adopted solar minimum heating rates from Strobel (2008a) to compare with the SHE model results. Further studies should be done using a 3D model as the amounts of CH$_4$ and CO in the atmosphere are better constrained.

We first obtained a converged solution for density and temperature versus radial distance with the fluid/kinetic approach using the HS collision model with the DSMC simulation. To compare the effects of using the HS, HS-LB, VHS and VHS-LB collision models on $\varphi$, $n(r)$, $T(r)$ and $u(r)$ within the DSMC model, the following lower boundary conditions were adopted from the converged fluid/kinetic (HS) model $r_{od} = 2836$ km where $\lambda_{od} = 12$,

469     $n_{od} = 2.9 \times 10^7$ cm$^3$ and $T_{od} = 85.5$ K. Using the conductivity for
470     the Maxwell molecules at $T_0$ (Strobel, 2008a), we obtained a
471     reference value for the HS cross section of $\sigma_0 = 9.0 \times 10^{-15}$ cm$^2$, see
472     Table 1 for the model parameters. The mean free paths at the
473     lower boundary for a collision for HS and VHS molecules are $l_c =$
474     $(\sqrt{2}n\sigma_0)^{-1}$ and $l_c = (<v_{th}>/n\sigma_0<v_{r0}>)$ respectively (Bird 1994), but
475     the values of $l_c$ and $Kn(r_{od})$ for this particular case are similar: ~27
476     km and ~0.1 respectively. While the resulting density profiles did
477     not significantly depend on the choice of the collision model for
478     the parameters $n_{od}$ and $\lambda_{od}$ given above (e.g. for all results $r_x \sim$
479     3900 km), the resulting temperature profiles and escape rates were
480     slightly different: e.g., the escape rates are 4.4, 5.1, 4.3 and 4.8
481     $\times 10^{25}$ s$^{-1}$ for the HS, HS-LB, VHS and VHS-LB models
482     respectively. Above the exobase, as it is seen in Fig. 2a the
483     translational temperature decreases faster than the rotational
484     temperature, and the perpendicular temperature decreases faster
485     than the radial temperature. At distances increasingly above the
486     exobase the atmosphere cools approximately adiabatically as
487     collisions become increasingly infrequent.
488     

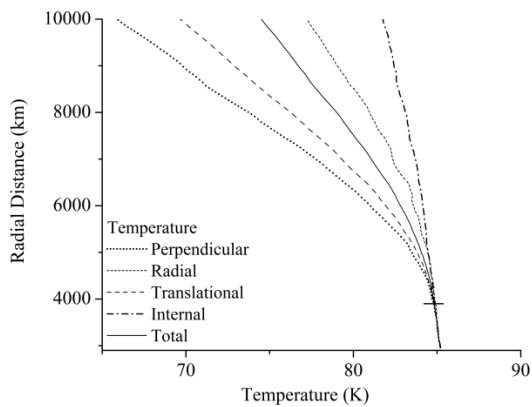

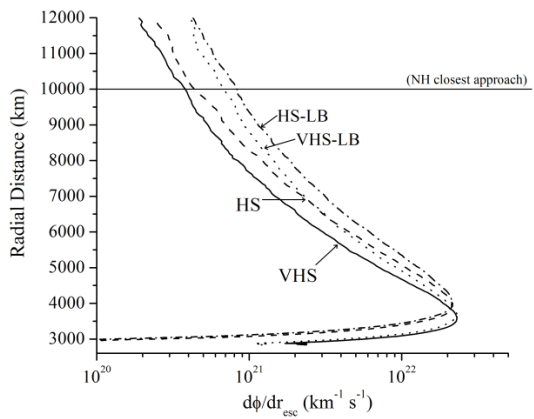

**Figure 2:** Results of test simulation described in text *Kn(rod)*=0.1 and $\beta(r) = 0$ above lower boundary: exobase altitude is 3900 km indicated by short horizontal curve in Fig. 2a.: (a) *T*(K) in VHS-LB model: short dashed curve perpendicular temperature, $T_{perp}$, dotted curve radial temperature, $T_r$, dashed dotted curve rotational temperature, $T_{rot}$, dashed curve translational temperature, $T_{trans} = (T_r + 2T_{perp})/3$, solid curve total temperature, $T = (3T_{trans} + 2T_{rot})/5$. (b) Production of escaping molecules, $d\varphi/dr_{esc}$ (km$^{-1}$s$^{-1}$) vs. *r*: HS (dashed curve), VHS (solid curve), HS-LB (dashed dotted curve),

VHS-LB (dotted curve). The New Horizons spacecraft distance of closest approach to Pluto will be 10000 km.

In the transition region of the atmosphere there is an altitude where it is most efficient for molecules to acquire escape trajectories. Below this altitude collisions inhibit escape and above there are too few collisions to produce escape trajectories. In the kinetic region we calculated the average number of escaping molecules produced in each radial cell, $d\varphi/dr_{esc}$, by noting the altitude at which molecules that eventually traverse the top of the simulation domain, $r_1 = 30000$ km, first attained an escape velocity. Molecules that later lose their escape velocity are dropped from the inventory, so the total escape rate is given by $\varphi = \int_{r_0}^{r_1} (d\varphi/dr)_{esc} dr$.

The peak in the escape rate production, Fig. 2b, occurs at the same altitude for the HS ($r \sim 4090$ km) and VHS ($r \sim 3680$ km) models with and without the internal degrees of freedom. The difference in the peak altitude is determined by the conductivity which differs between the VHS ($\kappa \, \alpha \, T$) and HS models ($\kappa \, \alpha \, T^{1/2}$). For the fluid/kinetic results discussed further below we used the VHS-LB model in order to have both $\kappa(T)$ and $C_p$ consistent with the fluid model, and to allow for rotational/ translation energy exchange.

**Table 2: SHE vs. fluid/kinetic**

| $\beta_0$ ($10^{-3}$ erg cm$^{-2}$ s$^{-1}$) | No heating | 1.7 | 1.5 |

|  | SHE* | fluid/kinetic~ | SHE* | fluid/kinetic~ |
|---|---|---|---|---|
| $r_x$ (km) | 2700 | 3900 | 3530 | 6200 |
| $n_x$ ($\times 10^5$ cm$^{-3}$) | 53 | 17 | 53 | 6.7 |
| $T_x$ (K) [$H_x$(100km)] | 48 [1.2] | 85 [4.5] | 65 [2.6] | 87 [12] |
| $u_x$ (m s$^{-1}$) | 1 | $5 \times 10^{-4}$ | 2 | 4 |
| $\lambda_x$ | 23 | 8.8 | 13 | 5.4 |
| $\varphi$ ($10^{25}$ s$^{-1}$) | 54 | 4.8 | 180 | 120 |
| $\varphi/\varphi_J$ | ~$10^7$ | 1.6 | ~$10^3$ | 2.0 |
| $\langle E\varphi \rangle / kT_0\varphi$ | 0 | 1.8 | 0 | 1.8 |
| $\langle E\varphi \rangle_J / kT_0\varphi_J$ | ~1.11 | 2.02 | ~1.53 | 2.12 |

Exobase values for the density, temperature, bulk velocity, escape rate φ and average energy carried off $\langle E\varphi \rangle$ including the corresponding values for the theoretical Jeans escape rate and energy flow rate ($\varphi_J$, $\langle E\varphi \rangle_J$) evaluated at the corresponding exobase distances: simulations performed for $\beta_0=0$ and solar minimum conditions for the SHE model and the fluid/kinetic (VHS-LB) results shown in Fig. 3. The lower boundary radial distance is $r_0 = 1450$ km where $n(r_0) = 4 \times 10^{12}$ cm$^{-3}$ ($Kn_0$~$10^{-6}$), $T(r_0) = 88.2$ K, $\lambda(r_0) = 23$ and $c_0 = 191$ m/s (sound speed). ~The exobase altitude $r_x$ is determined where $Kn = l_c/H = 1$, $l_c = \langle v_{th} \rangle / (n\sigma_0 \langle v_{r0} \rangle)$. Results are given in Table 2 from two cases

535    obtained using the combined fluid/kinetic simulation with the
536    VHS-LB collision model. In the first case no solar heating occurs
537    in the simulation region: i.e., $Q(r) = 0$ above $r_0$ so that $\beta_0=0$. Next
538    we assume approximate solar minimum conditions where the net
539    heating/cooling above $r_0$ is such that $\beta_0= 1.5 \times 10^{-3}$ erg cm$^{-2}$ s$^{-1}$
540    which is similar to the value used in the SHE model $\beta_0= 1.7 \times 10^{-3}$
541    erg cm$^{-2}$ s$^{-1}$ (Strobel, 2008a). The value of $\beta_0$ was obtained by
542    using a fixed solar UV and EUV heating efficiency, $\varepsilon \sim 0.25$, with a
543    cut-off in the heating at an altitude where the heat deposited was
544    less than 1% of $\beta_0$ (Strobel, 2008a).
545        As seen in Table 2, the escape rate obtained for the $\beta_0=0$
546    case, $4.8 \times 10^{25}$ s$^{-1}$, is $\sim 1.6 \times \varphi_J$, where $\varphi_J$ is the Jeans rate for $T_x= 85$
547    K and the heat flow out is $\sim 1.4 \times \langle E\varphi \rangle_J$ both evaluated at the
548    exobase, $r_x = 3900$ km where $Kn = 1$. These results differ
549    significantly from those obtained in the SHE model (e.g., $r_x = 2700$
550    km, $n_x \sim 5.3 \times 10^6$ cm$^{-3}$, $T_x = 48$ K and $\lambda_x \sim 23$; Strobel, 2008a)
551    indicative of the very different atmospheric profiles as seen in Fig.
552    3a. The change in temperature with increasing $r$ is seen to fall off
553    much faster in the SHE solution consistent with an overestimate in
554    the adiabatic cooling due to the overestimate in the escape rate.
555    Even though the SHE model has a much larger Jeans parameter at
556    the exobase, the escape rate is an order of magnitude larger than
557    that obtained with the fluid/kinetic model.

558

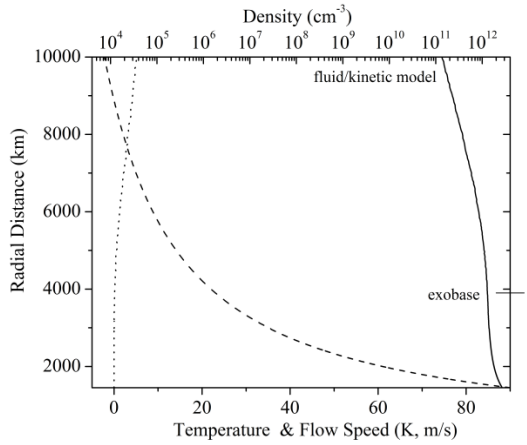

559

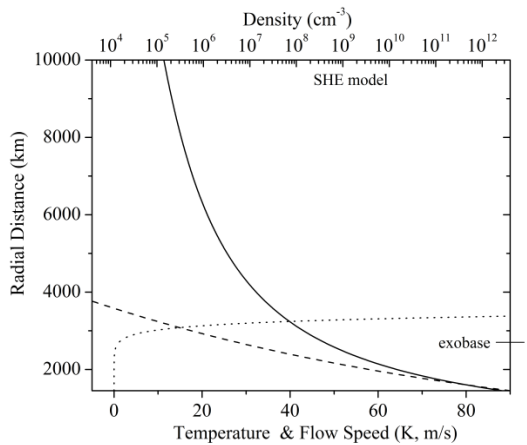

560

561

**Figure 3:** $n$(cm$^{-3}$) (top axis), $T$(K) and $u$(m/s) (bottom axis) vs. radial distance: Comparison of fluid/kinetic (**a**) $n$(dashed curves), $T$(solid curves) and $u$(dotted curves) to SHE model results (**b**) from Strobel (2008a) for no heating above $r_0$ ($\beta_0 = 0$). The exobase distance is indicated by the solid curve on right axis: 3900 km fluid/kinetic model and 2700 km SHE model. The New Horizons spacecraft distance of closest approach to Pluto will be 10000 km.

Numerically solving the fluid equations for $\beta_0 \neq 0$ is *very sensitive* to the choice of the input escape parameters, especially the energy carried away by escape, $\langle E\varphi \rangle$. Therefore, an initial solution was achieved by incrementally adding in a small fraction of the heating rate and solving the fluid equations iteratively, but assuming Jeans escape at the exobase, $Kn=1$. Having achieved a converged solution in this manner, we used the calculated $n(r_{od})$ and $T(r_{od})$ evaluated at $Kn(r_{od}) = 0.1$ as the starting point for the first DSMC iteration. When the solar minimum heating is included the Jeans parameter at the lower boundary for the kinetic model for the converged result was $\lambda(r_{od}) \sim 9$. The SHE result taken from Strobel (2008a) is obtained by solving Eq. 1b and 1c using assumed values of $\varphi$ and $(dT/dr)_0$, to find a solution with $n, T \rightarrow 0$ and zero total energy flux as $r \rightarrow \infty$.

As seen in Table 2, the resulting escape rate for the solar minimum case was $1.2 \times 10^{27}$ s$^{-1}$ with $r_x \sim 6200$ km, $n_x \sim 7 \times 10^5$ cm$^{-3}$, $T_x \sim 87$ K and $\lambda_x \sim 5$. Although, the escape rate is fortuitously close to the SHE result, $1.8 \times 10^{27}$ s$^{-1}$, the structure of the exobase region for the SHE model is *very* different: $r_x \sim 3530$ km, $n_x \sim 5.3 \times 10^6$ cm$^{-3}$, $T_x \sim 65$ K and $\lambda_x \sim 13$. Therefore, although the SHE escape rate was suggested to be $>10^3 \times \varphi_J$, based on the temperature and density at the exobase obtained here the escape rate is $2.0 \times \varphi_J$

592    and the energy flux rate is $1.7 \times \langle E\varphi \rangle_J$. The size of this

593    enhancement to the Jeans rate is similar to that found earlier

594    (Tucker and Johnson, 2009). Ignoring the effect on Charon, this

595    rate is also ~84% of the energy-limited escape rate,

596    $(4\pi r_0^2 \beta_0)/(\lambda_0 k T_0)$, often used in exoplanet studies (Lammer et al.,

597    2009).

598

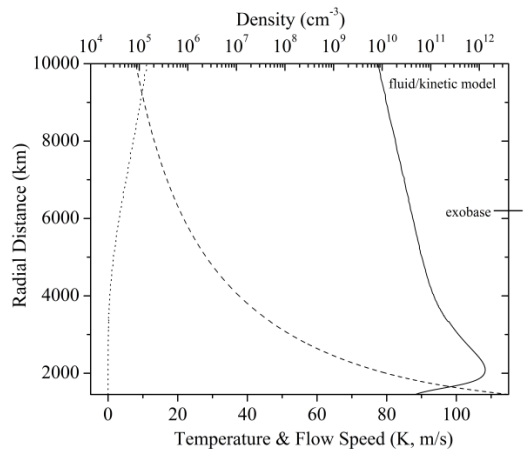

599

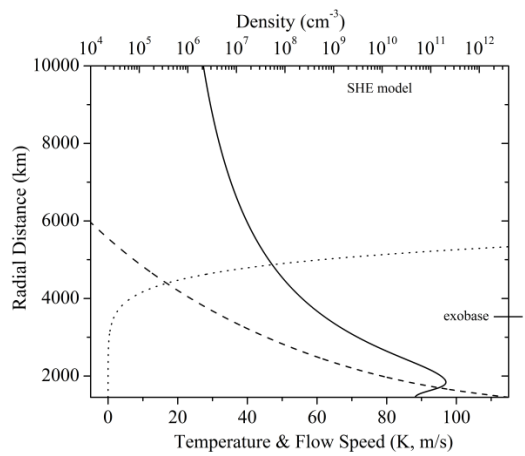

600

601

**Figure 4:** $n$(cm$^{-3}$) (top axis), $T$(K) and $u$(m/s) (bottom axis) vs. radial distance: Comparison of fluid/kinetic (**a**) $n$(dashed curves), $T$(solid curves) and $u$(dotted curves) to SHE model results (**b**) from Strobel (2008a) for solar minimum heating conditions above $r_0$. The exobase distance is indicated by the solid curve on right axis: 6200 km fluid/kinetic model and 3530 km SHE model. The New Horizons spacecraft distance of closest approach to Pluto will be 10000 km.

## 5  Effect of Charon on Escape

We examine here whether or not these results have implications for Charon, which has ½ the diameter and $1/10$ mass of Pluto. At an orbital distance of 19500 km the Hill sphere radius about Charon is at a radial distance of 12700 km from Pluto, where the atmospheric density is ~$10^5$ cm$^{-3}$ for the solar minimum case, Fig. 4a. Ignoring here any tidal effect on Pluto's lower atmosphere we estimated whether the effect of Charon's gravity on the molecular trajectories would significantly affect the escape rate. Therefore, we used the DSMC results for solar minimum conditions to perform free molecular flow (FMF) simulations in which molecules move under the influence of gravity from both Pluto and Charon but without inter-molecular collisions. Charon is assumed to have a circular orbit about Pluto and the molecular

625  trajectories are tracked in a 3D region from 10000 km to 30000 km
626  about Pluto. The FMF simulation is begun by emitting molecules
627  at $r = 10000$ km with radial velocities obtained from the
628  fluid/kinetic model for solar minimum heating conditions.
629  Escaping molecules are either emitted initially with a speed above
630  the escape speed or they gain an escape speed under the
631  gravitational influence of Charon. After several Charon orbits we
632  obtained a steady-state morphology of the gas density in the Pluto-
633  Charon system and the integrated escape rate produced versus
634  radial distance, as shown in Fig. 5(a, b).

635

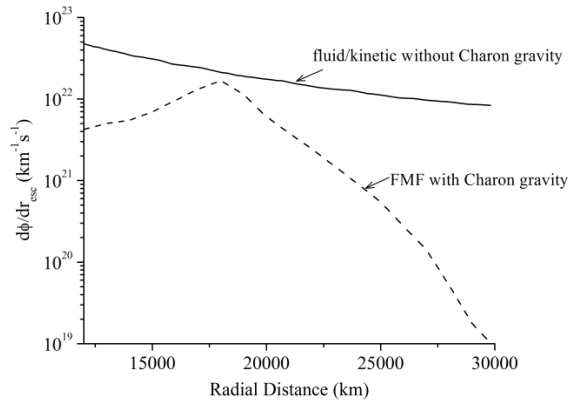

636

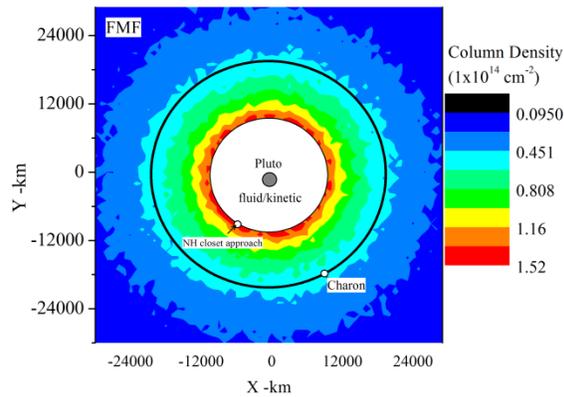

**Figure 5:** (a) Production of escaping molecules $d\varphi/dr_{esc}$ (km$^{-1}$s$^{-1}$) vs. $r$. DSMC results for solar minimum heating conditions without including Charon's gravitation influence (solid curve), as in Fig. 2b, compared to the fluid/kinetic result coupled to a FMF model that includes Charon's gravitational effect (dashed curve). (b) Contour plot of column density in the z plane Charon's northern and southern hemispheres in FMF simulations including Charon's gravitational influence: black curve represents Charon's orbit. Particles are emitted radially with corresponding speeds according to the local distribution at 10000 km. The New Horizons (NH) spacecraft distance of closest approach to Pluto will be 10000 km.

For solar minimum heating conditions we estimated that molecules from Pluto's atmosphere impinge upon Charon at a rate of $10^{25}$ s$^{-1}$. With Charon's surface temperature of ~50K this would be equivalent to the deposition of a monolayer of molecules over 4

Charon orbits or 8 x10$^{-3}$μm/yr. We also found that Charon has only a small effect on the escape rate from Pluto's atmosphere. Charon does not trap many molecules from Pluto's expanded atmosphere, but rather perturbs the molecules trajectories accelerating them to or decelerating them from escaping the system. The above conclusions were determined by performing FMF simulations with and without the gravitational influence of Charon in which we found that the escape rate decreased by 3 % in simulations that included Charon. This is opposite to the change in the escape rate that would have occurred had we used the energy to reach the Hill sphere of Charon as the escape criterion in the fluid/ kinetic simulations. A contour plot of the averaged total column densities over the north and south hemispheres when including Charon's gravitational influence is shown in Fig. 5b. The relevance of Charon is likely more significant at solar maximum heating conditions and close to perihelion.

## 6  CONCLUSIONS

Hydrodynamic models have often been applied to atmospheres in the solar system and to exoplanet atmospheres in order to estimate escape and the concomitant adiabatic cooling of the upper atmosphere (e.g., McNutt, 1989; Krasnopolsky, 1999; Tian and Owen, 2005; Strobel, 2008a, b; Yelle, 2004; Tian, 2009; Murray-Clay et al., 2009). Unless the Jeans parameter is < 2 well below the exobase (Kn<<1), this procedure can give incorrect atmospheric properties as compared to the fluid/kinetic combined approach described here. The difficulties with using a continuum model of thermal escape are twofold; how to, without prior knowledge, define density, temperature and energy flow at infinity, and how to properly define thermal conduction in the exosphere. The Fourier heat flux used to solve Eq. 2b becomes invalid near the exobase as discussed earlier (e.g., Johnson, 2010; Volkov et al., 2011a, b). Here we use a combined fluid/kinetic model that explicitly incorporates how heat conduction powers escape without requiring any assumptions about the macroscopic properties of the atmosphere at infinity. The hydrodynamic equations are solved below the exobase, and the kinetic model is continued above where the flow is essentially non-equilibrium. Such a procedure is relevant not only to Pluto but to the evolution of atmospheres on terrestrial bodies including recently discovered hot, rocky exoplanet atmospheres.

For over a few decades, hydrodynamic models have been used to conclude that Pluto's atmosphere is lost by a process called slow hydrodynamic escape (e.g., Krasnopolsky, 1999; McNutt, 1989; Strobel, 2008). We reconsidered escape from Pluto using the fluid/kinetic model and found that for the two cases considered, thermally-driven escape occurs at a rate within a factor of two of

700   the Jeans rate for the temperature determined in the combined

701   model. That is, for a lower boundary, $r_0$, in Pluto's atmosphere

702   where $\lambda(r_0) \sim 23$ and $Kn(r_0) \sim 10^{-6}$, and with all of the heat

703   deposited below $r_0$ (i.e., $\beta_0 = 0$), we obtain an escape rate $\varphi \sim 4.8$

704   $\times 10^{25}$ $N_2$ $s^{-1}$. For the derived exobase temperature, $T_x = 85$ K, this is

705   ~1.6 times the Jeans rate ($\varphi_J \sim 3.0 \times 10^{25}$ $N_2$ $s^{-1}$) and 1.4 times the

706   Jeans energy flux ($\langle E\varphi \rangle_J \sim 7.40 \times 10^{11}$ ergs $s^{-1}$). Furthermore we

707   find that each escaping molecule carries off an energy $\sim 2kT_0$ as

708   seen in Table 2, and not 0 as assumed in the SHE model. It is

709   interesting to note that for the same lower boundary conditions, if

710   one assumed the atmosphere was hydrostatic, the Jeans rate would

711   be $\varphi_J \sim 5.6 \times 10^{25}$ $N_2$ $s^{-1}$ and $\langle E\varphi \rangle_J \sim 1.4 \times 10^{12}$ ergs $s^{-1}$. Therefore,

712   these simulations indicate escape is similar in nature to Jeans

713   escape, but to get the correct *exobase temperature and density*

714   needed to make a Jeans estimate, a kinetic model should be applied

715   in the non-equilibrium region of the atmosphere.

716   We also simulated Pluto's atmosphere for solar minimum

717   conditions above $r_0$. For the same density and temperature at $r_0$,

718   with a similar solar minimum heating rate to that used in Strobel

719   (2008a), we obtain $\varphi \sim 1.2 \times 10^{27}$ $N_2$ $s^{-1}$. At the derived $T_x = 87$ K

720   from the model this is ~2.0 times the Jeans rate ($\varphi_J \sim 6.0 \times 10^{26}$ $N_2$

721   $s^{-1}$) and ~1.7 times the Jeans energy flux ($\langle E\varphi \rangle_J \sim 1.6 \times 10^{13}$ ergs $s^{-}$

[1]). As seen in Table 2, the escape rate for solar minimum conditions is fortuitously close to that obtained in Strobel (2008a) of $1.8 \times 10^{27}$ $N_2$ $s^{-1}$, but the total energy flux into the lower boundary leads to a very different atmospheric structure in the exobase region. Although this energy flux is a small fraction of the energy flux added above $r_0$ due to solar heating, $\langle E\varphi \rangle_{r_0} / (4\pi r_0^2 \beta_0) = 6.8 \times 10^{-2}$, it influences density and temperature gradients below the heating peak where $\beta(r) \to 0$ as $r \to r_0$. Furthermore the flow fields are radically different. This may be illustrated by comparing the relative magnitudes of the static, $p$, and dynamic, $\tfrac{1}{2}mnu^2$, pressures of each model. For the radial distance examined, 1450 km – 10000 km, we find the total pressure profile of the fluid/kinetic model monotonically decreases and is dominated by the static pressure. On the other hand, the SHE model experiences a minimum in the dynamic pressure at 5700 km with the region below dominated by the static pressure and the region above dominated by the dynamic pressure. The sum of the dynamic and static pressures in the SHE model also exceeds that of the fluid/kinetic model above 8200 km, suggesting that despite their similar flux rates the two models may be observationally distinguishable in their determination of Pluto's interaction with the solar wind.

We show here that starting at a small $Kn(r_0)$ in Pluto's atmosphere, a combined fluid/kinetic model can lead to reliable energy and molecule escape rates both for no heating and solar minimum heating conditions in the region above 1450 km. It is also clear from the fluid and DSMC results in the overlap region, that for this range of Jeans parameters a fluid model *can* obtain accurate temperatures densities and gas velocities with similar heat fluxes up to the exobase. But this is the case *only* if the φ and $\langle E\varphi \rangle_{r_0}$ used are equal to that obtained from a kinetic simulation of the exobase region. In fact, the total energy flux through the system cannot be determined independently for finite $Kn(r_0)$ using a fluid calculation, because it depends on the flow in the non-equilibrium region of the exosphere. Numerical methods have been used to solve the hydrodynamic model using a Jeans type escape and energy flux at or near the exobase (Chamberlain, 1961; Yelle, 2004; Gruzinov, 2011). However, these models require assumed values for $n$ and $T$ at the upper boundary.

We have treated Pluto's atmosphere using a single species, $N_2$, throughout the simulation region and have found an enhanced Jeans rate like that found earlier (Tucker and Johnson, 2009). Although minor species, with very different masses, will separate from the $N_2$ profile in the region of escape (e.g., Tucker and Johnson, 2009), CO should roughly track the $N_2$ profile described here. Since the solar activity during the observations of Greaves et al. (2011) in 2009/10 was close to that used for our assumed solar minimum conditions, the discovery of CO at altitudes ~4500km might not be surprising based on the atmospheric structure found here and does not require understanding the interaction of the extended atmosphere with the solar wind. Based on an assumed mixing ratio of ~0.05% the CO tangential column density at 4500 km would be ~ $6\times10^{11}$ CO cm$^{-2}$, but the temperature at this altitude is 91 K as opposed to 50 K suggested by the observations.

Solar maximum conditions are expected to occur in 2013, so that the New Horizon encounter with Pluto and Charon in 2015 will occur somewhere between solar maximum and minimum conditions. At a distance from the sun of 33 AU and assuming the same heating efficiency and cooling process, this results in $\beta_0$ ~1.7 times that used here (~$2.5\times10^{-3}$ erg cm$^{-2}$ s$^{-1}$). Therefore, accurate simulations of the atmospheric density at the encounter distance 10000 km from Pluto, and the atmospheric structure and the escape rates expected during the encounter will require the use of a fluid/kinetic model such as that described here. Such calculations are in progress for a multispecies atmosphere.

**Acknowledgements:** We thank D. Strobel for the solar minimum heating rates, L. Young for information on Pluto's atmosphere, and

790    R. Yelle and A. Gruzinov for helpful comments on atmospheric

791    escape. We also note the loss of two pioneers in the field whose

792    papers we heavily relied upon, D. Hunten and J. Elliot. This

793    research was supported by the NASA Planetary Atmospheres

794    Program and the NSF Astronomy Program.
41

883     Young, E. F. et al., 2008. Vertical Structure in Pluto's Atmosphere
884             from the 2006 June 12 Stellar Occultation. Astrophys. J.
885             136, pp. 1757-1769.